\documentclass[prl,aps,twocolumn,showpacs,preprintnumbers,amsmath,amssymb]{revtex4}
\usepackage{bm}
\usepackage{epsfig}

\begin{document}

\preprint{}

\title{Magnetic-field-induced Fermi surface reconstruction in Na$_{0.5}$CoO$_2$}

\author{L. Balicas,$^1$ M. Abdel-Jawad,$^2$ N. E. Hussey,$^{2}$ F. C. Chou,$^3$ and P. A. Lee$^4$}

\affiliation{$^1$National High Magnetic Field Laboratory, Florida
State University, Tallahassee-FL 32306, USA} \affiliation{$^2$H.
H. Wills Physics Laboratory, University of Bristol, Tyndall
Avenue, Bristol BS8 1TL, UK} \affiliation{$^3$Center for Materials
Science and Engineering, Massachusetts Institute of Technology,
Cambridge, Massachusetts 02139, USA} \affiliation{$^4$Department
of Physics, Massachusetts Institute of Technology, Cambridge,
Massachusetts 02139, USA}

\date{\today}%
\begin{abstract}
We have performed electrical transport measurements at low
temperatures and high magnetic fields in Na$_{0.5}$CoO$_2$ single
crystals. Shubnikov de Haas oscillations were observed for two
frequencies $F_1 \simeq 150$ and $F_2 \simeq 40$ T corresponding
respectively to 1 and .25\% of the area of the orthorhombic
Brillouin zone. These small Fermi surface (FS) pockets indicate
that most of the original FS vanishes at the charge ordering (CO)
transition. Furthermore, in-plane magnetic fields strongly
suppress the CO state. For fields rotating within the conducting
planes we observe angular magnetoresistance oscillations (AMRO),
whose periodicity changes from two- to six-fold at the transition,
suggesting that a reconstructed hexagonal FS emerges at a field of
about 40~T.

\end{abstract}

\pacs{71.18.+y, 72.15.Gd, 71.30.+h} \maketitle

The discovery of superconductivity in hydrated
 Na$_x$CoO$_2$ has stimulated intense interest in this material.\cite{takada}
 The conductive CoO$_2$ layers can be regarded as an
electron-doped correlated $S=1/2$ triangular network of frustrated
Co spins.  These electronically active triangular planes of edge
sharing CoO$_6$ octahedra are separated by Na and hydration layers
that act not only as spacers, leading to electronic
two-dimensionality, but also as charge reservoirs
\cite{takada,foossc}. While many workers have drawn analogy to the
high $T_c$ cuprates, the Na$_x$CoO$_2$ system is surely of great
interest in its own right as one of the few examples of a strongly
correlated material with a frustrated lattice where the carrier
concentration can be tuned continuously.

The rich phase diagram of the non-hydrated Na$_x$CoO$_2$ system as
a function of the Na content $x$ \cite{fooprl} reveals a
succession of ground states, from a paramagnetic metal at $x$ =
0.3 through a charge-ordered (CO) insulator at $x = 1/2$, a
``Curie-Weiss metal" for $x \simeq 0.70$, and finally a
magnetically ordered state for $x > 0.75$. At $x = 0.5 $, detailed
electron \cite{zandbergen} and neutron \cite{huang} diffraction
measurements reveal that Na orders in an orthorhombic
superstructure that is commensurate with the underlying lattice.
Notably, the Na ordering sets in above room temperature though
resistivity measurements only show a sharp rise below 53~K,
leading to an insulating ground state.\cite{fooprl}  The nature of
this ground state is unclear. It is presumably induced by the Na
ordering, but the low ordering temperature and the sharp onset
suggests that it is a nontrivial collective state rather than a
band insulator which results from an enlargement of the unit cell
due to Na ordering.  Photoemission results are not available for
the insulating samples, but work on $x = 0.6$ and 0.7 reveals the
existence of a large hole-like pocket, with an average radius $k_F
\sim 0.65 \pm 0.1$ \AA$^{-1}$, centered around the $\Gamma$
point.\cite{Hasan} These studies also reveal a very narrow
bandwidth $W \sim 100$ meV, confirming the strongly correlated
nature of the cobaltates for $x = 0.6$ and 0.7.

\begin{figure}[htb]
\begin{center}
\epsfig{file=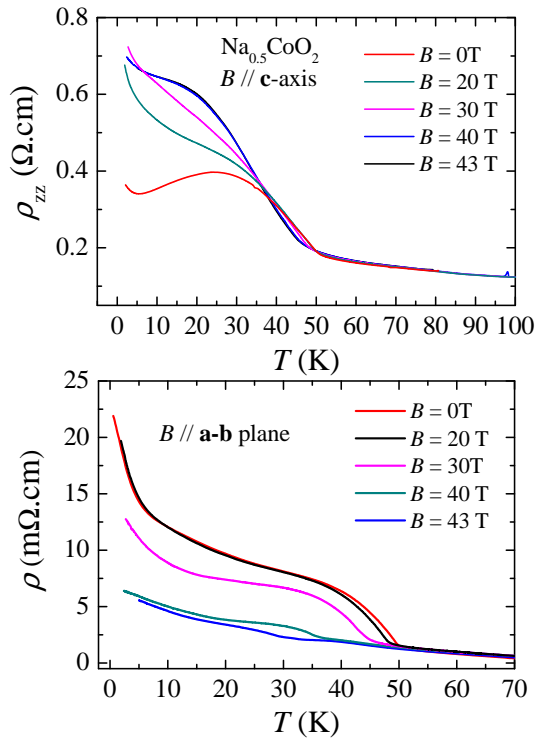, width=7 cm} \caption{Upper panel:
Inter-plane resistivity $\rho_{zz}$ as a function of temperature
$T$ for a Na$_{0.5}$CoO$_2$ single crystal and for several values
of the external magnetic field $B$ applied along the inter-plane
direction. The sharp increase in resistivity corresponds to a
transition towards a charge ordered state. Lower panel: In-plane
resistivity $\rho$ as a function of $T$ and for several values of
$B$ applied along the conducting planes. Notice the progressive
suppression of the CO-state for $B>20$ T. }
\end{center}
\end{figure}
\begin{figure}[htb]
\begin{center}
\epsfig{file=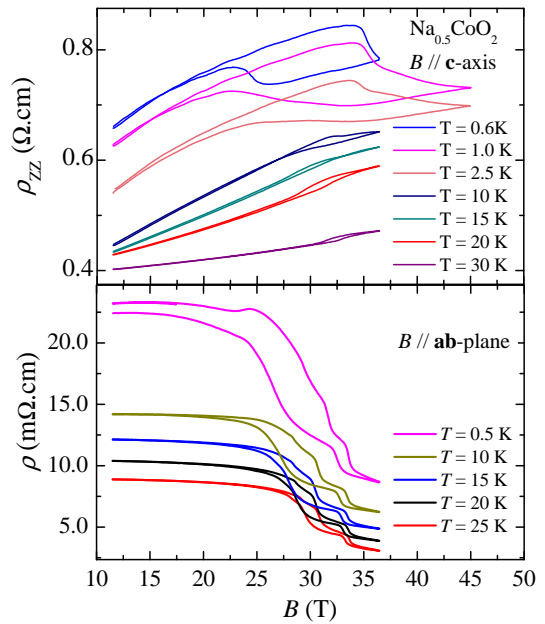, width=7 cm} \caption{Upper panel:
Inter-plane resistivity $\rho_{zz}$ as a function of magnetic
field $B$ for a Na$_{0.5}$CoO$_2$ single crystal and for several
values of temperature. Lower Panel: In-plane resistivity $\rho$ as
a function of $B$ applied along the planes, for a second
Na$_{0.5}$CoO$_2$ single crystal and for several values of $T$.
Notice the sharp decrease of $\rho$ for $B >25$ T.}
\end{center}
\end{figure}
The key question is what happened to the large hole pocket in the
insulating ground state.  To gain some insight into this question,
we have performed a detailed electrical transport study in
Na$_{0.5}$CoO$_2$ at high fields $B$ and low temperatures $T$. We
find that the CO-state observed below $T_{\text{co}} = 53$ K can
be suppressed by large in-plane magnetic fields, but not by fields
applied along the inter-plane direction. For $B$ rotating within
the conductive CoO$_2$ layers we observe angular magnetoresistance
oscillations of essentially two-fold periodicity consistent with
the reported orthorhombic crystallographic symmetry of
Na$_{0.5}$CoO$_2$ \cite{huang}. As $B$ increases (i.e., as the
CO-state is suppressed) however, a new 6-fold periodicity emerges
indicating the stabilization of a hexagonal FS as reported by the
ARPES measurements \cite{Hasan}. This observation suggests on the
one hand, that the Na superstructure defines the geometry of the
FS at low temperatures, and on the other, that the charge order in
the conducting plane is suppressed by high in-plane fields. At low
temperatures Shubnikov de Haas oscillations (SdH) of very small
frequencies are observed for $B
\parallel c$-axis, indicating that almost the entire FS reported
for $x = 0.6$ and 0.7 \cite{Hasan} disappears below
$T_{\text{co}}$ for $x=0.5$. Our results strongly indicate that
the charge ordering involves the coupling with the Na order and
involves the large hole pocket rather than the small pockets near
the $K$ points as proposed in Refs. \cite{singhprb,pickettprb,
Johannesprl}.

Single crystals of Na$_{0.75}$CoO$_2$ were grown using the
floating-zone technique. By using an electrochemical
de-intercalation procedure, samples were produced with a nominal
Na concentration  $x = 0.5$, as confirmed by Electron Microprobe
Analysis. Details of the crystal growth process, the
electrochemical de-intercalation technique, and the
characterization of the resulting samples are discussed in detail
in Ref. \cite{chou}.  Electrical transport measurements were
performed using the standard four-terminal technique in a two-axis
rotating sample holder inserted in $^3$He cryostat. High magnetic
fields up to $B$ = 45 T were provided by the hybrid magnet at the
NHMFL in Tallahassee.
\begin{figure}[htb]
\begin{center}
\epsfig{file=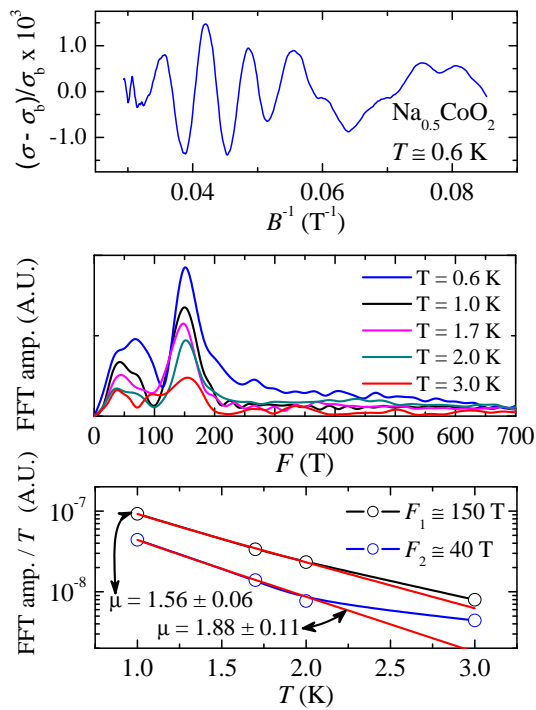, width=7 cm} \caption{Upper panel: The
Shubnikov de Hass signal, defined as $(\sigma - \sigma_b)/
\sigma_b$, where $\sigma = 1/ \rho_{zz}$ and $\sigma_b = 1/
\rho_b$ with $\rho_b$ being the background resistance, as a
function of the inverse of the magnetic field $B^{-1}$. Middle
panel: The FFT of the SdH signal for several temperatures. Two
peaks are observed at $F_1 \simeq 150$ T and $F_2 \simeq 40 $ T,
respectively. Lower panel: The amplitude of both peaks in the FFT
signal, normalized respect to the temperature $T$, and as a
function of $T$. Red lines are fits to the Lifshitz-Kosevich
expression $x/sinhx$, from which we extract the effective masses
$\mu$. }
\end{center}
\end{figure}

The upper panel of Fig.1 shows the $T$-dependence of the
inter-plane resistivity $\rho_{zz}$ of a Na$_{0.5}$CoO$_2$ single
crystal for several values of the external field $B$ applied along
the inter-plane c-axis. The abrupt change in slope corresponds to
the onset of the charge-ordering at $T_{\text{co}}= 53$ K. $B$
applied along the c-axis has very little effect on
$T_{\text{co}}$. Indeed, the absolute value of $\rho_{zz}$
actually increases at lower $T$s, presumably due to field-induced
renormalization of the c-axis dispersion (due to Landau
quantization, see below). The lower panel of Fig. 1 shows the
in-plane resistivity $\rho$($T$) for $B$ applied along an in-plane
direction. Note how $T_{\text{co}}$ as well as the low $T$
resistivity is markedly suppressed by $B>20$ T applied along the
CoO$_2$ planes. The suppression of a CO-state by an external field
is not surprising since it has been shown that a charge-density
wave (CDW) can be destabilized by an external field surpassing the
Pauli limit \cite{graf, ross}. In our case, the anisotropy of the
Pauli critical field could be attributed to the anisotropy of the
Land\'{e} $g$ factor \cite{choususcept}, since the susceptibility
is highly anisotropic. By analogy to superconductors, the Pauli
critical field in a one-dimensional CDW is estimated to be $ B_P =
\Delta_0/ \left( \sqrt{2} gS \mu_B \right)$.\cite{ross} The
optical gap has been measured to be 13.64~meV or
158~K.\cite{timusk}  Interpreting this as $2\Delta_0$, and
assuming $g = 2$, $S = {1 \over 2}$, we estimate $B_P \approx
83$~T.  This is somewhat larger than the observed field of 40~T,
but not unreasonable given the uncertainty concerning $g$ and
$\Delta_0$ and the applicability of the one dimensional model.

Figure 2 shows both $\rho_{zz}$ as a function of $B\parallel$
c-axis (upper panel) as well as $\rho$ as a function of $B\perp$
c-axis (lower panel) for several values of the temperature $T$.
Notice how inter-plane fields in excess of 25 T have a modest but
quite hysteretic effect on $\rho_{zz}$, suggesting a partial
destabilization of the CO-state. By contrast, fields applied along
the conducting planes lead to a significant (factor of 3)
reduction in $\rho$. The effect of $B$ is clearly far more
pronounced for in-plane fields. The steps seen in $\rho$ at higher
$B$ could correspond to metastable configurations of the CO state.
\begin{figure}
\begin{center}
\epsfig{file=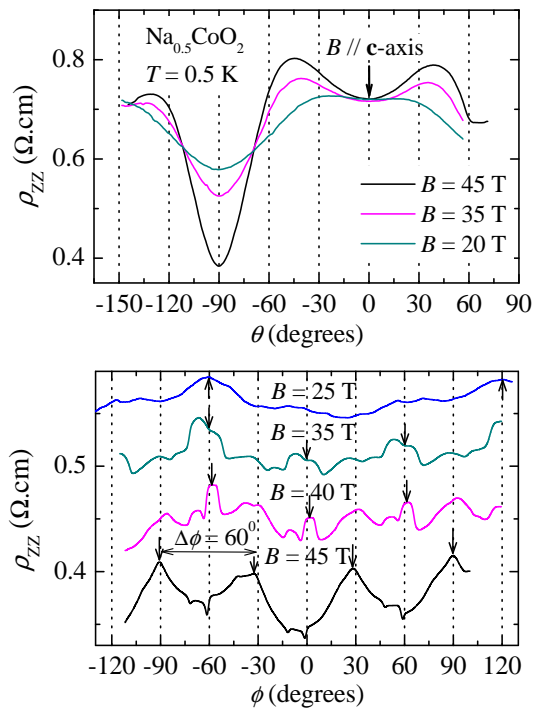, width=7 cm} \caption{Upper panel:
Inter-plane resistivity $\rho_{zz}$ as a function of the angle
$\theta$ between the external magnetic field $B$ and the
inter-plane c-axis and for several values of the external field.
The pronounced drop in $\rho$ for $B \parallel $ a-b plane is due
to the partial suppression of the CO-state at this orientation.
Lower panel: Angular magnetoresistance oscillations for field
applied along the conducting planes. Notice how the nearly two-
fold symmetric behavior progressively disappears in favor of a
6-fold one as the CO state is suppressed.}
\end{center}
\end{figure}
The most significant feature of the $\rho_{zz}$ data is
undoubtedly the appearance of SdH oscillations at low $T$ as seen
in the upper panel of Fig. 2. The top panel in Fig. 3 shows an
example of the isolated SdH signal defined as $(\sigma -
\sigma_b)/ \sigma_b $, where $\sigma$ is the inverse of
$\rho_{zz}$ and $\sigma_b$ is the inverse of the background
resistivity as a function of inverse field $B^{-1}$ at $T = 0.6$
K. Note that only a few oscillations are observed over a broad
range of $B$. The middle panel of Fig. 3 shows the amplitude of
the FFT signal as a function of the SdH frequency $F$ for several
values of $T$. A main peak with $F_1 \simeq 150$ T is clearly
observed in the temperature evolution of the FFT signal with a
second weaker peak appearing at $F_2 \simeq 40$ T. Using the
Onsager relation $F = A(hc/4 \pi^2 e)$, where $A$ is the FS
cross-sectional area perpendicular to $B$, and the orthorhombic
lattice parameters reported in Ref. \cite{huang}, these
frequencies correspond to pockets with cross-sectional areas
occupying only 1 and .25\% of the area of the orthorhombic
Brillouin zone, respectively. Note that this is only 0.25 and
0.0625\% of the undistorted hexagonal Brillouin zone. Of course
SdH only tells us the area of each pocket and not the number of
pockets. Nevertheless, a change of Fermi surface topology and a
dramatic reduction in the carrier density is consistent with Hall
measurements which show a sign change at the transition
temperature and an increase of the magnitude of $R_H$ by a factor
of 250. \cite{fooprl}. For $B \parallel c$-axis no SdH
oscillations were observed in the hysteretic region, probably due
to the absence of a well defined coherent FS around the
transition.

In the Lifshitz-Kosevich \cite{LK} (LK) formalism, the ratio of
each harmonic $p$ of the oscillatory component of the conductivity
$\widetilde{\sigma} = (\sigma - \sigma_b)$ to the monotonically
varying background $\sigma_b$ has the form:
\begin{equation}\label{LK}
    \frac{\widetilde{\sigma}}{\sigma_b} \propto
    \frac{(-1)^p}{p^{3/2}}R_{T,p}R_{D,p}R_{s,p}\left[2\pi p\left(\frac{F}{B}-\gamma\right)\pm\frac{\pi}{4}\right]
\end{equation}
where $R_{T,p} = (\alpha p \mu T/B)/\sinh(\alpha p \mu T/B)$,
$R_{D,p}= \exp[-\alpha p \mu T_D/B]$ and $R_{s,p}$ are the
temperature, Dingle and spin damping factors, respectively
\cite{LK}. In the above expressions $\alpha = 2\pi^2 k_B m_e/e
\hbar$, $\mu$ is the quasiparticle's effective cyclotron mass in
relative units of the free electron mass $m_e$, and $T_D$ is the
so-called Dingle temperature $T_D= \hbar/2 \pi k_B \tau$ that
provides an accurate value for the quasiparticles relaxation time
$\tau$. Thus, in order to obtain the effective masses $\mu$ for
both orbits $F_1$ and $F_2$ we plot in the lower panel of Fig. 3
the amplitude of the FFT signal normalized respect to the
temperature $T$. The red lines are fits to the temperature damping
factor $R_{T,p}$ from which we obtain $\mu_1 = 1.6 \pm 0.1$ for
$F_1$ and $\mu_2 = 1.9 \pm 0.1$ for $F_2$, respectively. These
effective masses are light if we recall that an effective mass of
order 70 $m_e$ was extracted for $x = 0.6$ and $x = 0.7$ from
ARPES and in specific heat measurements.  The light mass however
is quantitatively consistent with the reduction of the Sommerfeld
coefficient for $x=0.5$, obtained by extrapolating the heat
capacity $C/T$ to $T=0$ \cite{huang}. This is an important point
since it indicates the absence of additional Fermi surfaces of
heavier masses that might not have been detected in the present
study. Finally, in order to estimate the quality of our samples,
we performed a semi-logarithmic Dingle plot, obtaining the
surprisingly low value $T_D \simeq 3.3 \pm 0.5 $ K. From $T_D$ we
obtain $\tau = (3 \pm 0.5) \times 10^{-13} $ s, corresponding to
$\omega_c \tau \simeq 0.85 \pm 0.13 $ at 25 T and a mean free path
$l = v_F \tau = \hbar (A/\pi)^{1/2} \tau / \mu \sim 145$ \AA. This
implies $k_Fl = 9.7$ where $k_F = (A/\pi)^{1/2}$ is the radius of
the pockets.  Thus the observation of SdH oscillations indicate
that the crystal is sufficiently clean and that the residual Fermi
pocket should be metallic. However, from Fig.~1, lower panel, we
see that the resistivity is insulating-like, due to the loss of
charge carriers when the gap opens. One explanation is that the
bulk resistivity is not intrinsic, but dominated by scattering
from CDW domain boundaries in some complicated way. Alternatively,
the SdH signal may originate from a small part of the sample where
the Na is particularly well ordered.

Angular magnetoresistance oscillations have been proven to be a
powerful way of extracting the 3D geometry of the FS in overdoped
cuprates \cite{nigelnature}. We are currently applying these
techniques to Na$_x$CoO$_2$ and the current results obtained for
$x = 0.5$ are displayed in Fig. 4 which shows the dependence of
the inter-plane magnetoresistance $\rho_{zz}$ on the angle
$\theta$ between $B$ and the inter-plane c-axis, at $T = 0.6$ K
and for several values of $B$.  A pronounced dip is observed when
the field is aligned along the in-plane direction, due to the
destabilizaton of the CO-state. Two peaks are observed at $\pm
45^{\circ}$ that are most likely associated with genuine Yamaji
effects whose angular position depends on both the FS size and the
complexity of the c-axis dispersion\cite{nigelnature}. The lower
panel of Fig. 4 shows $\rho_{zz}$ as a function of the azimuthal
angle $\phi$ between $B$ and an in-plane axis. At lower fields,
$\rho_{zz}(\phi)$ is essentially two-fold, with some small poorly
defined structures at intermediate angles. Notice that this
structure cannot result from a cylindrical FS of circular
cross-section, since $k_F$, $v_F$ and $\tau$ would remain uniform
across its perimeter and consequently $\rho_{zz}(\phi)$ would
remain constant. The observed two-fold periodicity reflects, in
fact, the symmetry of the orthorhombic structure resulting from
the Na ordering. A similar effect was reported in the tetragonal
Tl$_2$Ba$_2$CuO$_6$ compound \cite{nigelprl}, where
$\rho_{zz}(\phi)$ displays a 4-fold periodicity due to the 4-fold
modulation induced on $k_F$, $v_F$, and $\tau$  by the structure
of the CuO$_2$ planes. In our case, the two-fold dependence of
$\rho_{zz}(\phi)$ progressively disappears as $B$ increases (with
the emergence, at the CO to metallic transition, of field
dependent fine structures) and stabilizes into a 6-fold one at
higher fields. This periodicity is consistent with both the
triangular symmetry of the CoO$_2$ planes and with the hexagonal
FS measured by ARPES \cite{Hasan}. It is remarkable that once the
CO state is suppressed, the resulting metallic state seems
oblivious to the Na order.

In summary, we have presented a detailed electrical transport
study on Na$_{0.5}$CoO$_2$, observing both Shubnikov-de Haas
oscillations and two-fold angular magnetoresistance oscillations
in its charge-ordered state. Both the frequency of the SdH
oscillations and the periodicity of the in-plane AMRO at low
fields indicate that the Na ordering introduces profound effects
on the electronic structure of this compound. Moreover, a
relatively modest in-plane magnetic fields is found to suppress
the charge ordered gap leading to a correlated metal. At high
fields, an in-plane AMRO displaying a 6-fold modulation emerges.
These facts indicate that the charge order is a delicate one, more
akin to a CDW, and consistent with the small optical
gap.\cite{timusk} Clearly, the interplay between Na ordering and
charge ordering in these strongly correlated materials remains a
challenging problem.

This work was performed at the NHMFL which is supported by NSF
through NSF-DMR-0084173, the State of Florida and DOE. One of us
(LB) acknowledges support from the NHMFL in-house research
program. FCC acknowledges support from the MR-SEC Program of the
National Science Foundation under award number DMR-02-13282 and
from DOE under grant number DE-FG02-04ER46134. PAL acknowledges
NSF award number DMR-0201069.

\end{document}